\documentclass[12pt,aps]{revtex4-1}

\usepackage{amsmath}  
\usepackage{amsfonts} 
\usepackage{graphicx} 
\usepackage{relsize}
\usepackage[symbol]{footmisc}
\DeclareMathOperator{\arctanh}{arctanh}

\begin{document}


\title{Cosmological horizons}

\author{Michele Re Fiorentin}
\email{michele.refiorentin@polito.it} 
\altaffiliation{Polytechnic University of Turin - Department of Applied Science and Technology} 
\author{Stefano Re Fiorentin} 
\email{s.refiorentin@gmail.com} 
\affiliation{  }



\begin{abstract}
The recently unveiled deep-field images from the James Webb Space Telescope have renewed interest in what we can and cannot see of the universe. 
Answering these questions requires understanding the so-called ‘‘cosmological horizons’’ and the ‘‘Hubble sphere’’.
Here we review the topic of cosmological horizons in a form that university physics teachers can use in their lessons, using the latest data about the so-called standard ‘‘Lambda Cold Dark Matter’’ ($\Lambda$CDM) model. 
Graphical representations are plotted in terms of both conformal and proper coordinates as an aid to understanding the propagation of light in the expanding universe at various epochs.

\end{abstract}

\maketitle 

\renewcommand{\thefootnote}{\fnsymbol{footnote}}

\section{Introduction} 
\label{Sec1}

Amber Straughn, an astrophysicist at NASA's Goddard Space Flight Center in Maryland, said during the live event of presentation of the first James Webb's images \cite{2022 Crane}: ‘‘Today, for the first time, we're seeing brand new stars that were previously completely hidden from our view.’’
This might suggest that with technological evolution we will be able to see virtually everything in the universe. Actually, this statement should be understood as limited to the so-called \textit{observable universe}, bounded by the \textit{particle horizon}. 
Even within the observable universe, two questions can be asked: (i) Is it possible to observe galaxies that, due to the expansion of the universe, were receding at superluminal speeds when they emitted their light? (ii) Waiting even for infinite time, will it be possible in the future to see the entire universe?
Answers to these questions require the knowledge of the so-called \textit{cosmological horizons}, a topic that has already been the subject of several publications. 
The first comprehensive one, aimed at providing a clear definition and representation of horizons is Rindler's 1956 paper \cite{1956 Rindler}, which summarizes all previous work on the subject. However, it is obsolete today. 
Another noteworthy article is by Edward Harrison \cite{1991 Harrison} which, however, was written in 1991, well before the discovery of the accelerated expansion of the universe. 
The first educational article devoted to the topic of cosmological horizons belongs to the same period (1993) and was published in this same journal \cite{1993 Ellis} by George Francis Rayner Ellis and Tony Rothman. The treatment, however, was limited to a matter-only universe, which is no longer acceptable today.
More recent and up-to-date works on the topic are those by Berta Margalef-Bentabol, Juan Margalef-Bentabol and Jordi Cepa \cite{2012 Margalef}, \cite{2013 Margalef} and that by Valerio Faraoni in Chapter 3 of \textit{Cosmological and Black Hole Apparent Horizons} \cite{2015 Faraoni}.
However, none of these references provide calculations of the horizons with the current parameters of the $\Lambda$CDM model. This literature therefore cannot be used directly to calculate the path of photons emitted from different sources in the expanding universe and to evaluate how these sources are located with respect to the horizons.
It is worth mentioning that there have also been publications aimed at correcting misconceptions on the topic, such as the seminal ones by Tamara Davis \cite{2003 Davis, 2004 Davis, 2005 Lineweaver}  which, however, do not include a comprehensive treatment of cosmological horizons.

We will give here a didactic presentation of cosmological horizons in the case of a flat universe, in view of the fact that experimental evidence has shown that our universe does not deviate appreciably from flatness \cite{2020 Planck Collaboration}. 
The presentation is intended as a support for academics who want to teach the topic of cosmological horizons in introductory courses on Relativity or Cosmology.
Since all the formulas for the various quantities under consideration are derived in the text, students could draw the graphs themselves or devise new ones, using tools such as MATHEMATICA.
The first three sections are intended to provide the necessary background by summarizing known information.
In particular, in the next section we will briefly present Hubble's law, the Hubble radius and the past light cone. In the following section we will define cosmological horizons, while the third introductory section will be devoted to the derivation of the expansion law of the scale factor.

\section{The expanding universe}
\label{Sec2}

The starting point is the metric of the homogeneous and isotropic universe. This metric is known as the Friedmann–Lemaître–Robertson–Walker (FLRW) metric, that in spherical coordinates reads \cite{2019 Carroll}:
\begin{equation}
\label{eq:FLRW}
\mathbf{ds}^2=-c^2dt^2+a^2(t)\left[\frac{dr^2}{1-kr^2}+{r^2}\left(d\theta^2+{\sin^2\theta}\,d\varphi^2\right)\right],
\end{equation}
where $k$ is the curvature parameter.
We will be concerned with the case of a flat universe, where $k=0$. 
The dimensionless function $a(t)$, called the \textit{scale factor}, accounts for the expansion, while $r$ is called the \textit{comoving radial coordinate}. 
It is customary to assume that, at the present time $(t=t_0)$, $a(t_0)=1$, so that the comoving radial coordinate represents the proper radial distance at the present time.
At the generic time $t$, the proper radial distance is given by
\begin{equation}
\label{eq:Radial coordinate}
R(t)=a(t)r(t).
\end{equation}
The redshift $z$ is the experimental observable that allows direct evaluation of the scale factor at the time of the emission of the radiation we receive today by means of the relation \cite{2002 Coles}:
\begin{equation}
\label{eq:redshift}
1+z=\frac{1}{a}
\end{equation}

\noindent Considering a cosmic object moving in space with time-varying comoving radial coordinate $r(t)$ and taking the derivative of expression (\ref{eq:Radial coordinate}) with respect to time, we obtain for the physical velocity of the cosmic object
\begin{equation}
\label{eq:velocities}
v(t)=\frac{dR(t)}{dt}=r(t)\frac{da(t)}{dt}+a(t)\frac{dr}{dt}=v_{\scriptscriptstyle\mathrm{rec}}(t)+v_{\scriptscriptstyle\mathrm{pec}}(t),
\end{equation}
where we have introduced the \textit{peculiar velocity} of the cosmic object
\begin{equation}
\label{eq:peculiar}
v_{\scriptscriptstyle\mathrm{pec}}(t)\equiv{a(t)}\frac{dr}{dt},
\end{equation}
and the \textit{recession velocity} due to the expansion of space
\begin{equation}
\label{v_rec}
v_{\scriptscriptstyle\mathrm{rec}}(t)\equiv{r(t)}\frac{da(t)}{dt}=[a(t)\,r(t)]\left[\dfrac{1}{a(t)}\frac{da(t)}{dt}\right]=R(t)\,H(t).
\end{equation}
Here we have defined the function
\begin{equation}
\label{Hubble}
H(t)\equiv\frac{1}{a(t)}\frac{da(t)}{dt},
\end{equation}
called \textit{Hubble parameter}, whose value at the present time, denoted by $H_0\equiv{H(t_0)}$, is customarily (but improperly) called the \textit{Hubble constant}.
The relation (\ref{v_rec}) evaluated at the present time
\begin{equation}
\label{Hubble_law}
v_{\scriptscriptstyle\mathrm{rec}}(t_0)=H_0\,R(t_0)
\end{equation}
constitutes the so-called Hubble's law, experimentally found and published by Hubble in 1929 \cite {1929 Hubble}.
The expression (\ref{eq:velocities}) is useful for determining the velocity that a photon travelling in space along radial directions has with respect to us. In fact, knowing that the peculiar speed of light is always equal to $c$, this formula allows us to obtain the velocity relative to us of a photon that at time $t$ is at the comoving radial coordinate $r$ and moving radially toward us ($v_{\scriptscriptstyle\mathrm{pec}}=-c$):
\begin{equation}
\label{v_photon}
v_{\scriptscriptstyle\mathrm{phot}}(t)=r\frac{da(t)}{dt}-c=v_{\scriptscriptstyle\mathrm{rec}}(t)-c.
\end{equation}
We observe that if $v_{\scriptscriptstyle\mathrm{rec}}>c$, the photon's velocity becomes positive and thus pointing away from us. 
The comoving radial coordinate $r_{\scriptscriptstyle{H}}$ such that at time $t$ we have $r_{\scriptscriptstyle{H}}(t)\,da/dt=c$ is called \textit{comoving Hubble radius} at time $t$ and it delimits the \textit{Hubble sphere}:
\begin{equation}
\label{Hubble_radius}
r_{\scriptscriptstyle{H}}(t)\equiv{c}\left(\frac{da}{dt}\right)^{-1}.
\end{equation}
A photon emitted from a source that has a comoving radial coordinate greater than $r_{\scriptscriptstyle{H}}$ fails to approach us at the moment of emission. 
However, as we shall see, the Hubble sphere also expands with time  so that in some cases it comes to ‘‘encompass‘‘ the photon which then moves towards us.

As is well known, photons travel along null geodesics. If we consider photons moving along radial directions, from the FLRW metric (\ref{eq:FLRW}) with $k=0$ we can write that
\begin{equation}
\label{light}
a(t)dr={\pm}cdt.
\end{equation}
The positive sign applies to  photons travelling away from the origin, while the negative sign describes photons moving toward the origin.
Taking into account Eq. (\ref{eq:peculiar}), the relation (\ref{light}) reflects the fact that the peculiar speed of light is always equal to $c$.
By integrating the equation (\ref{light}) with the negative sign between the space-time point of emission and our space-time point of observation $(t=t_0,r=0)$, we obtain the comoving distance travelled by light from the point of emission to us. Denoting this distance by $r_{\scriptscriptstyle{LC}}(t_0;t)$ we get the definition 
\begin{equation}
\label{light_cone}
r_{\scriptscriptstyle{LC}}(t_0;t)\equiv\int_{t}^{t_0}\frac{cdt'}{a(t')}\qquad(t<t_0).
\end{equation}
The function $r_{\scriptscriptstyle{LC}}(t_0;t)$ for all $t<t_0$ describes \textit{our past light cone}, which represents the locus of space-time points from which the photons that reach us now were emitted.
If instead of present time $t_0$ we consider the generic time $t^*$, we obtain the past light cone relative to time $t^*$:
\begin{equation}
\label{light_cone_t*}
r_{\scriptscriptstyle{LC}}(t^*;t)\equiv\int_{t}^{t^*}\frac{cdt'}{a(t')}\qquad(t<t^*).
\end{equation}
\section{The horizons}
\label{Sec3}

Each past light cone has a maximum distance $r_{\scriptscriptstyle{LC}}(t^*;0)$.
This is the comoving radial distance travelled by a radiation (e.g. gravitational) emitted at time zero and received at time $t^*$.
It has been given the name of \textit{comoving particle horizon} and we denote it by $r_{\scriptscriptstyle{PH}}(t^*)$. 
Removing the superscript $^*$, the comoving particle horizon at time $t$ is therefore defined by
\begin{equation}
\label{particle_horizon}
r_{\scriptscriptstyle{PH}}(t)\equiv\int_{0}^{t}\frac{cdt'}{a(t')}.
\end{equation}
Of particular interest is the comoving particle horizon at present time, $r_{\scriptscriptstyle{PH}}(t_0)$: it represents the farthest comoving distance from which we can retrieve information from the past and constitutes the boundary between the region whose events have already been observed, and the region whose events cannot yet be observed.
We mention that sometimes the \textit{optical horizon} $r_{\scriptscriptstyle{OH}}(t_0)$ is introduced \cite{2015 Faraoni}. 
It is simply the comoving radial distance travelled by photons emitted when light has decoupled from matter. These photons constitute the Cosmic Microwave Background Radiation (CMBR) and originate from the maximum distance one can see in the electromagnetic domain. 
Indicating with $t_{ls}$ the time of emission of the CMBR, the optical horizon is defined by
\begin{equation}
\label{optical}
r_{\scriptscriptstyle{OH}}(t)\equiv\int_{t_{ls}}^{t}\frac{cdt'}{a(t')}\qquad(t>t_{\scriptscriptstyle{ls}}).
\end{equation}
The concept of particle horizon is more relevant than that of optical horizon, since there could be gravitational waves that bring us information about what happened before photon decoupling.

The light cone relative to an infinite time $r_{\scriptscriptstyle{LC}}(\infty;t)$ is of particular interest.
It constitutes the locus of the space-time points $(t,r)$ from which the radiation will reach $r=0$ at an infinite time. The relevance of this light cone lies in the fact that it constitutes a horizon because a photon emitted from a position beyond this limit can never reach $r=0$. It is therefore called the \textit{comoving event horizon} at time $t$, and we denote it by $r_{\scriptscriptstyle{EH}}(t)$. From Eq. (\ref{light_cone_t*}) we get:
\begin{equation}
\label{event_horizon}
r_{\scriptscriptstyle{EH}}(t)\equiv\int_{t}^{\infty}\frac{cdt'}{a(t')}.
\end{equation}
To fully understand the implications that horizons and the Hubble sphere have on the visibility of cosmic objects, it is helpful to plot them in space-time diagrams.
To this end it is first necessary to determine the time dependence of the scale factor.

\section{Evolution of the scale factor}
\label{Sec4}
We can obtain the equations for determining the time dependence of the scale factor by substituting the metric (\ref{eq:FLRW}) into the Einstein field equations.
The stress-energy tensor for the isotropic, homogeneous universe is that of a perfect fluid, given by \cite{2019 Carroll}
\begin{equation}
\label{eq:T_mu-nu}
T^{\mu\nu}=\left(\rho+\frac{p}{c^2}\right)u^{\mu}u^{\nu}+pg^{\mu\nu},
\end{equation}
$\rho$ and $p$ being the mass density and pressure, respectively, and $u^\mu$ the 4-velocity vector field of the fluid.
Due to the symmetry of the metric (\ref{eq:FLRW}), the ten field equations reduce to two, known as Friedmann Equations, that in the case of $k=0$ are \cite{2002 Coles}:
\begin{eqnarray}
\label{eq:Friedmann_1}
\left(\frac{1}{a}\frac{da}{dt}\right)^2&=&\frac{8\pi{G}}{3}\rho+\frac{1}{3}\Lambda{c^2},\\
\label{eq:Friedmann_2}
\left(\frac{1}{a}\frac{d^2a}{dt^2}\right)&=&-\frac{4\pi{G}}{3}\left(\rho+\frac{3p}{c^2}\right)+\frac{1}{3}\Lambda{c^2},
\end{eqnarray}
where $\Lambda$ is the so-called cosmological constant that appears in the field equations. 
Using the first equation, the second can be rewritten as
\begin{equation}
\label{eq:conservation}
\frac{d\rho}{dt}=-3H\left(\rho+\frac{p}{c^2}\right),
\end{equation}
which expresses the conservation of mass-energy, since it coincides with $\nabla_{\mu}{T^\mu}_0=0$ \cite{2019 Carroll}. 
If the fluid is a mixture of two or more non-interacting fluids, such an equation holds separately for each fluid. This assumption is crucial for the solution of the Friedmann equations and is generally accepted. Furthermore, it is assumed that each fluid has an equation of state of the type
\begin{equation}
\label{eq:eq_of_state}
p=w\rho{c^2}
\end{equation}
where $w$ is a constant that takes the values $0$ for matter (both barionic and dark matter) and $1/3$ for radiation.
Using previous assumptions and substituting Eq. (\ref{eq:eq_of_state}) into Eq. (\ref{eq:conservation}), we arrive at
\begin{alignat}{3}
\label{eq:rho_matter}
&\rho_m(t)\,&&=\rho_{m0}\;a(t)^{-3}\qquad &&\mathrm{for\;matter},\\
\label{eq:rho_radiation}
&\rho_r(t)\,&&=\,\rho_{r0}\;a(t)^{-4}\qquad &&\mathrm{for\;radiation},
\end{alignat}
where $\rho_{m0}$ and $\rho_{r0}$ are the matter density and radiation density at the present time, respectively.
We will assume in the following that matter dominates over radiation, an assumption that only fails in the first $\sim30\,000$ year, a period of time that represents about two millionths of the current lifetime of the universe.
For simplicity, we will omit the subscript $_m$ in the matter density in the following. 
Substituting Eq. (\ref{eq:rho_matter}) into Eq. (\ref{eq:Friedmann_1}) we get
\begin{equation}
\label{eq:Friedmann_3}
\left(\frac{1}{a}\frac{da}{dt}\right)^2=\frac{8\pi{G}}{3}\rho_0{a(t)^{-3}}+\frac{1}{3}\Lambda{c^2}.
\end{equation}
Inserting this relationship into Eq. (\ref{Hubble}) and evaluating at the present time yields:
\begin{equation}
\label{Closure_1}
H_0^2=\frac{8\pi{G}}{3}\rho_0+\frac{1}{3}\Lambda{c^2}.
\end{equation}
It is customary to define the following dimensionless constants:
\begin{equation}
\label{eq:Constants}
\Omega_m\equiv\frac{8\pi{G}\rho_0}{3H_0^2},\qquad\Omega_{\Lambda}\equiv\frac{\Lambda{c^2}}{3H_0^2},
\end{equation}
so that Eq. (\ref{Closure_1}) gives
\begin{equation}
\label{Closure_2}
\Omega_m+\Omega_{\Lambda}=1.
\end{equation}
According to results published by the Planck Collaboration in 2020 \cite{2020 Planck Collaboration}, the most accepted values for $\Omega_m$, $\Omega_{\Lambda}$ and $H_0$, are\cite{uncertainties}
\begin{equation}
\label{values}
\Omega_m=0.3111\pm{0.0056},\qquad{\Omega_\Lambda}=0.6889\pm{0.0056},\qquad{H_0}=67.66\pm{0.42}\;\mathrm{(km/s)/Mpc}. 
\end{equation}
Substituting the definitions (\ref{eq:Constants}) into Eq. (\ref{eq:Friedmann_3}) we obtain
\begin{equation}
\label{eq:Friedamnn_4}
\frac{1}{a}\frac{da}{dt}={\pm}H_0\sqrt{(1-\Omega_{\Lambda}){a^{-3}}+\Omega_{\Lambda}}\qquad(a\geq0).
\end{equation}
Depending on the sign, this expression describes a monotonically increasing or decreasing scale factor.
Since the scale factor is currently increasing, we must choose the plus sign.
Furthermore, by defining $x=a^{3/2}$ the equation  (\ref{eq:Friedamnn_4}) can be put in the form: 
\begin{equation}
\label{to_be_integrated}
\frac{dx}{\sqrt{(1-\Omega_\Lambda)/\Omega_\Lambda+x^2}}=\frac{3H_0\sqrt{\Omega_\Lambda}}{2}dt\qquad(x\geq0).
\end{equation}
Knowing that
\begin{equation}
\frac{1}{\sqrt{\alpha^2+x^2}}=\frac{d}{dx}\sinh^{-1}\left(\frac{x}{\alpha}\right),
\end{equation}
the equation (\ref{to_be_integrated}) can be easily integrated. 
The lower integration limit for time is zero, while the lower limit for $x$ depends on the value that the scale factor takes on at time zero. 
It is generally assumed to be zero, but actually it could also be other than zero, as in the case where an initial period of inflation is considered. In any case, it is a very small value on a cosmological scale, so we will conform to the custom of considering it equal to zero.
We thus arrive at the following expression:
\begin{equation}
\label{a(t)}
a(t)=\left(\frac{1-\Omega_\Lambda}{\Omega_\Lambda}\right)^{1/3}\sinh^{2/3}\left(\frac{3H_0\sqrt{\Omega_\Lambda}}{2}t\right)\qquad(t\geq0).
\end{equation}
We emphasize that this time dependence of the scale factor is relative to the $\Lambda$CDM cosmological model that best describes the universe we live in, which assumes that it is flat  and that the contribution from radiation, affecting only the very early stages of cosmic expansion, is negligible.

\section{The horizons in conformal coordinates}
\label{Sec5}

The representation of cosmological horizons becomes particularly simple if we introduce a new time variable, denoted by $\eta$, defined by the relation
\begin{equation}
\label{eta(t)}
\eta(t)\equiv\int_{0}^{t}\frac{dt'}{a(t')},
\end{equation}
so that
\begin{equation}
\label{conformal time}
d\eta=\frac{dt}{a(t)}.
\end{equation}
The simplification occurs because this last expression is the integrand that appears in the equations (\ref{light_cone}), (\ref{particle_horizon}) and (\ref{event_horizon}) that define our past light cone, the particle horizon and the event horizon, respectively.
As a function of the new time variable, the FLRW metric (\ref{eq:FLRW}) with $k=0$ reads
\begin{equation}
\label{eq:FLRW_2}
\mathbf{ds}^2=a^2(\eta)\left[-c^2d\eta^2+dr^2+{r^2}\left(d\theta^2+{\sin^2\theta}d\varphi^2\right)\right],
\end{equation}
which is a flat Minkowski metric except for the conformal factor $a^2(\eta)$; for this reason $\eta$ is called \textit{conformal time}.
We will refer to the conformal time and to the comoving radial distance as ‘‘conformal coordinates’’.
In order to evaluate the integrals in equations (\ref{light_cone}), (\ref{particle_horizon}) and (\ref{event_horizon}) once the expression (\ref{conformal time}) has been inserted, we need to determine the values of $\eta(t_0)$ (that we denote with $\eta_0$) and of $\eta(\infty)$. 
This last, as we will see, in the case of the $\Lambda$CDM model assumes a finite value that we denote with $\eta_{\scriptscriptstyle\mathrm{max}}$. 
We will calculate these values in the next section.
Substituting the expression (\ref{conformal time}) into the equations (\ref{light_cone}), (\ref{particle_horizon}) and (\ref{event_horizon}) we get respectively:
\begin{alignat}{4}
\label{light_conf}
&\mathrm{Past\;Light\;Cone\;at\;}\eta_0: \quad &&r_{\scriptscriptstyle{LC}}(\eta_0;\eta)&&=\int_{\eta}^{\eta_0}c{d\eta}=c(\eta_0-\eta)\qquad &&(0<\eta<\eta_0),\\
\label{particle_conf}
&\mathrm{Particle\;Horizon}: \quad && r_{\scriptscriptstyle{PH}}(\eta)&&=\int_{0}^{\eta}c{d\eta}=c\eta\qquad &&(0<\eta<\eta_{\scriptscriptstyle\mathrm{max}}),\\
\label{event_conf}
&\mathrm{Event\;Horizon}: \quad && r_{\scriptscriptstyle{EH}}(\eta)&&=\int_{\eta}^{\eta_{\scriptscriptstyle\mathrm{max}}}c{d\eta}=c(\eta_{\scriptscriptstyle\mathrm{max}}-\eta)\qquad &&(0<\eta<\eta_{\scriptscriptstyle\mathrm{max}}).
\end{alignat}
These are equations of $\pm45^{\circ}$ straight lines in the conformal coordinates $(r,\eta)$ if we express $r$ in Glyr and $\eta$ in Gyr and adopt the same lengths for the units of Glyr and Gyr.

\section{Conformal diagram}
\label{Sec6}
To plot the horizons in conformal coordinates we need to calculate the values of $\eta_0$ and $\eta_{\scriptscriptstyle\mathrm{max}}$, as required by the equations (\ref{light_conf}) and (\ref{event_conf}). 
To this end, it is necessary to determine the function $\eta(t)$.
Substituting the expression (\ref{a(t)}) into Eq. (\ref{eta(t)}) we get
\begin{equation}
\label{eta}
\eta(t)=\frac{1}{\xi}\left(\frac{\Omega_\Lambda}{1-\Omega_\Lambda}\right)^{1/3}\int_{0}^{\xi{t}}\frac{dx}{\sinh^{2/3}(x)},
\end{equation}
having defined the constant
\begin{equation}
\label{xi}
\xi\equiv\frac{3H_0\sqrt{\Omega_\Lambda}}{2}=(8.609\pm0.064)\cdot10^{-2}\;\mathrm{Gyr}^{-1}.
\end{equation}
The integral that appears in Eq. (\ref{eta}) involves the Gaussian hypergeometric function  $_{\scriptscriptstyle{2}}{F}_{\scriptscriptstyle{1}}(a,b;c;z)$ \cite{hypergeometric}:
\begin{equation}
\label{hypergeometric}
\int\frac{dx}{\sinh^{2/3}(x)}=(-1)^{5/6}\cosh(x)\;_{\scriptscriptstyle{2}}{F}_{\scriptscriptstyle{1}}\left(\frac{1}{2},\frac{5}{6};\frac{3}{2};\cosh ^2(x)\right).
\end{equation}
In order to compute $\eta_0$ we derive the value of $\xi{t_0}$ from the equation (\ref{a(t)}) using the condition $a(t_0)=1$:
\begin{equation}
\label{xi_t_0_calc}
\xi{t_0}=\mathrm{arcsinh}{\sqrt{\frac{\Omega_\Lambda}{1-\Omega_\Lambda}}}=\arctanh\sqrt{\Omega_\Lambda}=1.188\pm{0.011}.
\end{equation}
Knowing the value of $\xi$ from Eq. (\ref{xi}), we get for the current age of the universe the value $t_0=13.80\pm0.10$ Gyr.
Substituting the value (\ref{xi_t_0_calc}) into the upper integration limit of the equation (\ref{eta}) and using the expression (\ref{hypergeometric}), we obtain $\eta_0=47.10\pm0.39\;\mathrm{Gyr}$.

\noindent In order to determine $\eta_\mathrm{max}$, we take into account that
\begin{equation}
\label{int_max}
\int_{0}^{\infty}\frac{dx}{\sinh^{2/3}(x)}=\frac{9}{\sqrt{\pi}}\,\mathlarger{\Gamma}\left(\frac{7}{6}\right)\mathlarger{\Gamma}\left(\frac{4}{3}\right)\simeq4.2065...
\end{equation}
Using this in Eq. (\ref{eta}) together with the other constants, we obtain $\eta_\mathrm{max}=63.69\pm{0.49}$ Gyr.
It is useful, for future purposes, to study the behaviour of $\eta(t)$ for both small and large values of $\xi{t}$.
In the first case, in the integrand of equation (\ref{eta}) we can replace $\sinh (x)$ with its series expansion near $x=0$, obtaining at first order:
\begin{equation}
\label{small_t_values}
\eta(t)\simeq\frac{3}{\xi}\left(\frac{\Omega_\Lambda}{1-\Omega_\Lambda}\right)^{1/3}({\xi}t)^{1/3}\qquad(\xi{t}\ll{1}).
\end{equation}
For large values of $\xi{t}$ it is appropriate to rewrite Eq. (\ref{eta}) in the form
\begin{equation}
\label{large_t_values}
\eta(t)=\eta_\mathrm{max}-\frac{1}{\xi}\left(\frac{\Omega_\Lambda}{1-\Omega_\Lambda}\right)^{1/3}\int_{\xi{t}}^{\infty}\frac{dx}{\sinh^{2/3}(x)}.
\end{equation}
The integral on the right-hand side, for large values of $x$, can be calculated by substituting $\sinh(x)$ with $e^{x}/2$, obtaining:
\begin{equation}
\label{large_t_values_2}
\eta(t)\simeq{\eta_\mathrm{max}}-\frac{3}{2^{\scriptscriptstyle{1/3}}\xi}\left(\frac{\Omega_\Lambda}{1-\Omega_\Lambda}\right)^{1/3}e^{-\frac{2}{3}\xi{t}}\qquad(\xi{t}\gg1).
\end{equation}
This expression shows that for large values of cosmic time, conformal time slows to a halt.
The function $\eta(t)$ is shown in Figure 1 on logarithmic scales. 
\begin{figure}[ht]
\includegraphics[width=.6\linewidth]{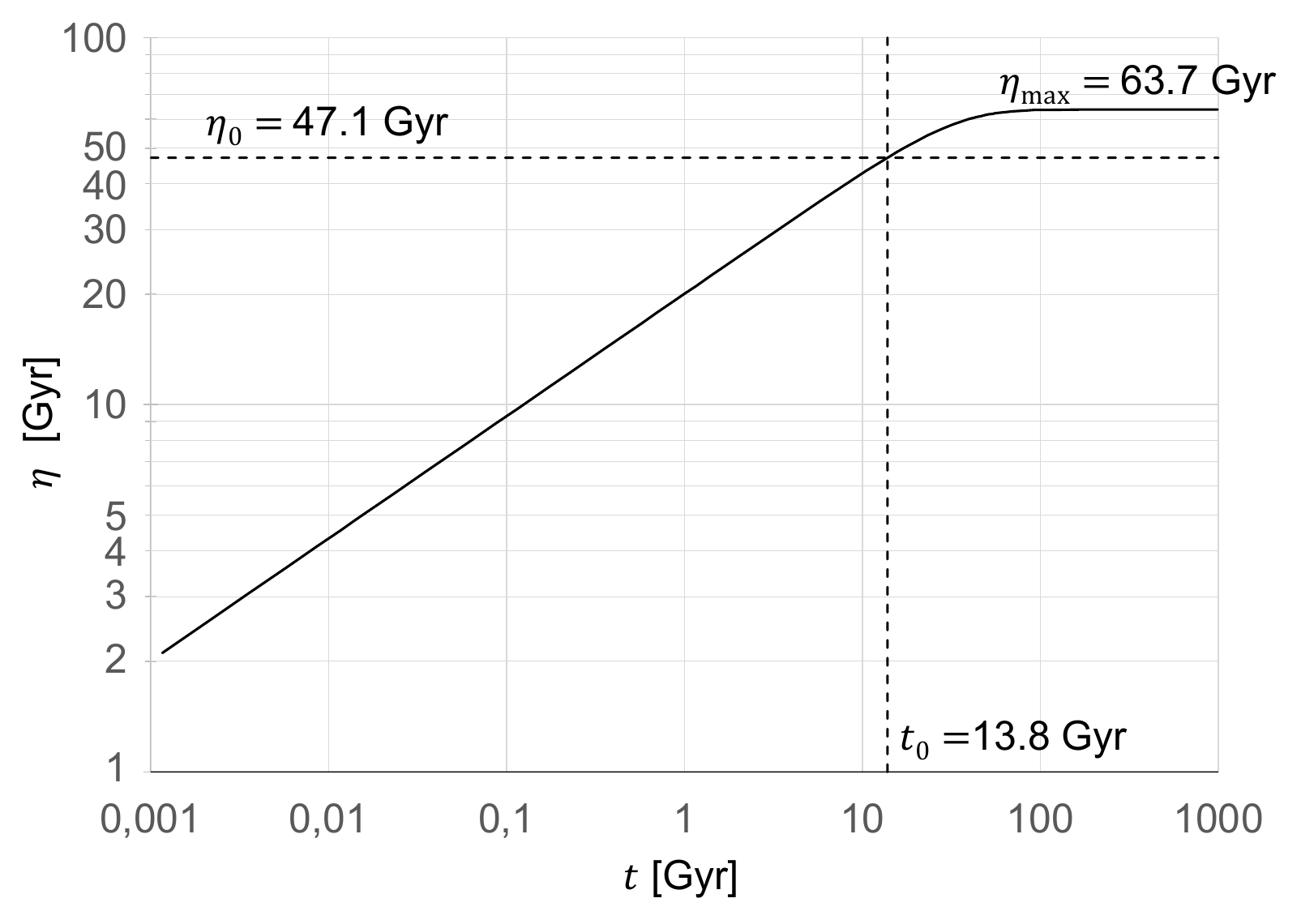}
\caption{Representation of conformal time $\eta$ as a function of cosmic time $t$}
\label{fig1}
\end{figure}
We are now ready to graph our past light cone and the two horizons according to the equations (\ref{light_conf}), (\ref{particle_conf}) and (\ref{event_conf}), plotting the conformal time vs. the comoving radial distance.
In this conformal representation the world lines of comoving cosmic objects are vertical straight lines.
To represent the Hubble sphere we use equation (\ref{Hubble_radius}) substituting in it the expression (\ref{a(t)}), thus obtaining
\begin{equation}
\label{rH_vs_t}
r_{\scriptscriptstyle{H}}(t)=\frac{3c}{2\xi}\left(\frac{\Omega_\Lambda}{1-\Omega_\Lambda}\right)^{1/3}\frac{\sinh^{1/3}\left(\xi{t}\right)}{\cosh\left(\xi{t}\right)}.
\end{equation}
To get $r_{\scriptscriptstyle{H}}(\eta)$ we would have to invert the equation (\ref{eta}) to get $t(\eta)$ to be substituted in Eq. (\ref{rH_vs_t}). Unfortunately, this step can only be performed numerically \cite{nintegration}. 
However, it is possible to obtain approximate expressions for both small and large values of $\xi{t}$.
In the first case, restricting ourselves to the first-order term of the series expansion of the right-hand side of Eq. (\ref{rH_vs_t}) in terms of $\xi{t}$ in the proximity of $\xi{t}=0$,  we get
\begin{equation}
\label{rH_vs_t_small_values_t}
r_{\scriptscriptstyle{H}}(t)\simeq\frac{3c}{2\xi}\left(\frac{\Omega_\Lambda}{1-\Omega_\Lambda}\right)^{1/3}(\xi{t})^{1/3}\qquad(\xi{t}\ll{1}).
\end{equation}
Using the expression (\ref{small_t_values}) we can then write the Hubble radius in terms of the conformal time as
\begin{equation}
\label{rH_vs_t_small_values_eta}
r_{\scriptscriptstyle{H}}(\eta)\simeq\frac{1}{2}c\,\eta\qquad(\eta\ll{1}).
\end{equation}
This is a straight line of slope $2$ in the plane $(r,c\eta)$.
Instead, for large values of $\xi{t}$ we can adopt for the expression (\ref{rH_vs_t}) the approximation
\begin{equation}
\label{rH_vs_t_large_values_t}
r_{\scriptscriptstyle{H}}(t)\simeq\frac{3c}{2^{\scriptscriptstyle{1/3}}\xi}\left(\frac{\Omega_\Lambda}{1-\Omega_\Lambda}\right)^{1/3}e^{-\frac{2}{3}\xi{t}}\qquad(\xi{t}\gg{1}).
\end{equation}
Then thanks to Eq. (\ref{large_t_values_2}) we can finally write $r_{\scriptscriptstyle{H}}$ as a function of $\eta$ :
\begin{equation}
\label{rH_vs_t_large_values_eta}
r_{\scriptscriptstyle{H}}(\eta)\simeq{c}\,(\eta_\mathrm{max}-\eta)\qquad(\eta\gg{1}).
\end{equation}
This is exactly the expression of the event horizon (\ref{event_horizon}): at large conformal times the Hubble sphere becomes tangent to the event horizon.
We will see later that this is because the physical velocity of photons proceeding along the event horizon (which is a null geodesic) tends progressively to zero, and thus the event horizon tends to the Hubble sphere.
It can be shown numerically that everywhere else (that is, for $\eta<\eta_\mathrm{max}$) relation  $r_{\scriptscriptstyle{EH}}(\eta)>r_{\scriptscriptstyle{H}}(\eta)$ results.

\noindent While the straight line described by Eq. (\ref{rH_vs_t_small_values_eta}) diverges from $r=0$, the straight line (\ref{rH_vs_t_large_values_eta}) converges to $r=0$. This means that the conformal Hubble radius must reach a maximum value. Unfortunately, it is not possible to determine it in closed form, but it can be found numerically that its value is $(r_{\scriptscriptstyle{H}})_\mathrm{max}\simeq16.52$ Glyr at conformal time $\eta=39.25$ Gyr (cosmic time $t=7.65$ Gyr).

All the curves are shown in Figure 2, which comprehensively illustrates the dynamics of each light emission event, thus making it clear what structures we can see today, will see in the future, and will never see, regardless of the resolving power of telescopes.
\begin{figure}[ht]
\includegraphics[width=.8\linewidth]{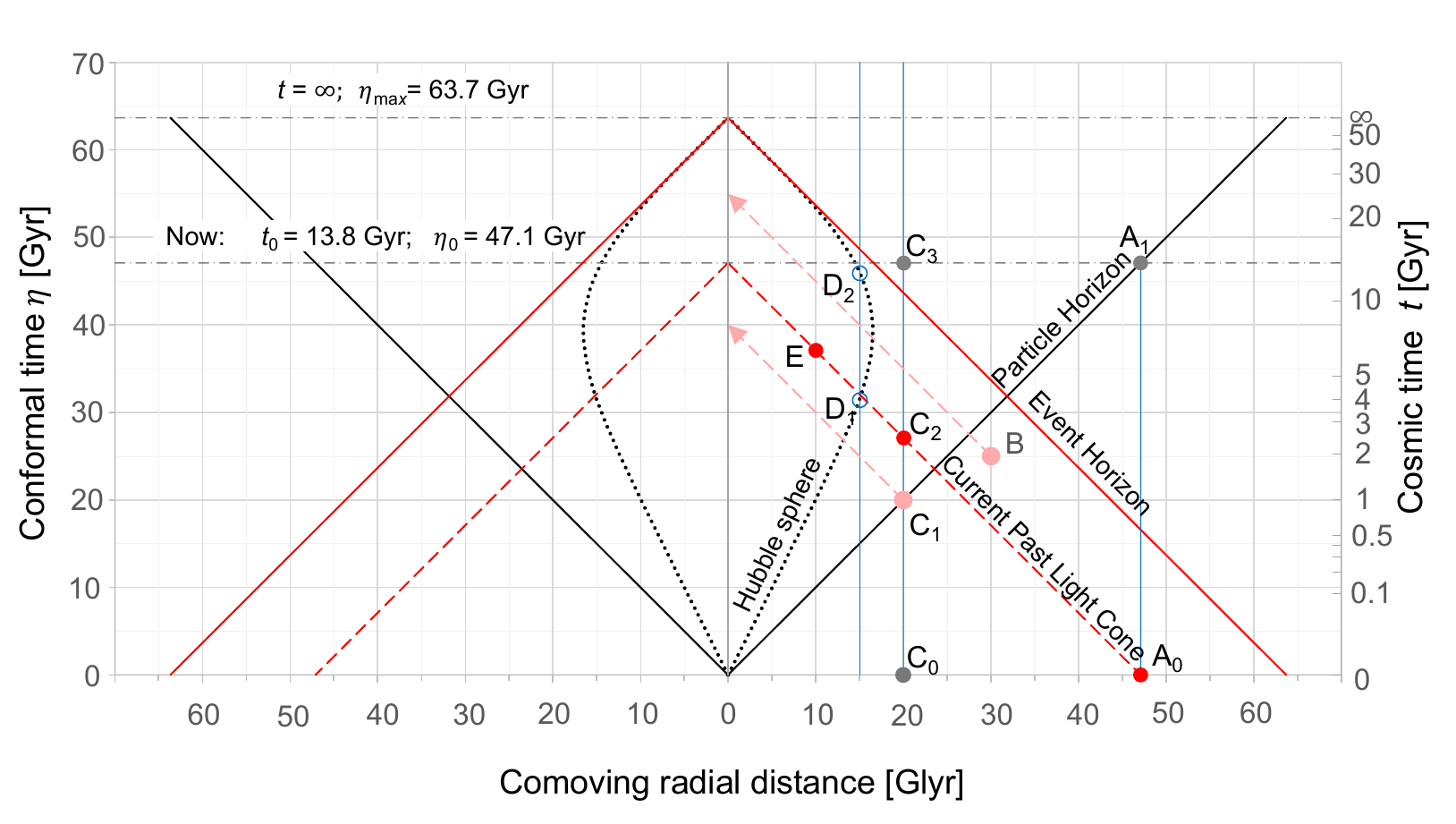}
\caption{Graph of past light cone at $t_0$, particle horizon, event horizon and Hubble sphere in conformal coordinates, together with the paths to the origin of photons emitted in events $A_0$, $B$, $C_0$, $C_1$, $C_2$ and $E$. Events $D_1$ and $D_2$ mark, respectively, the entry and exit from the Hubble sphere of the cosmic object placed at the comoving radial distance $r=15.0$ Glyr.}
\label{fig2}
\end{figure}
We begin by considering the $A_0$ event placed on our past light cone near time zero (or, for convenience, at the time of the surface of last scattering $t_{\scriptscriptstyle{ls}}\simeq380\,000$ years: it could be a spot in the CMBR with a slightly different temperature).
Such an emission event occurred at the comoving radial coordinate $r\simeq{c}\,\eta_0=47.10\pm0.39$ Glyr, well outside the Hubble sphere, whose radius goes to zero for $t\rightarrow0$.
So the photons emitted in $A_0$ toward the origin had a receding (positive) physical velocity relative to us. 
In particular, Eq. (\ref{v_rec}) with $t=t_{\scriptscriptstyle\mathrm{ls}}$ gives $v_{\scriptscriptstyle\mathrm{rec}}(t_{\scriptscriptstyle\mathrm{ls}})\simeq65\;c{\,}$, a highly superluminal. However, as can be seen from the Figure 2, in comoving coordinates the velocity was of approach (negative). This is because in conformal coordinates the expansion velocity is absent and thus the total velocity coincides with the peculiar velocity, which for light is always equal to $\pm{c}$. 
In fact, based on Eq.s (\ref{light}) and (\ref{conformal time}), in conformal coordinates the velocity of light always results, $v_{\scriptscriptstyle\mathrm{pec}}=dr/d\eta=\pm{c}$, so there cannot be superluminal velocities.
Therefore, in these coordinates, photons emitted in $A_0$ approached the Hubble sphere, which they reached at $\eta\simeq31.91$ Gyr ($t\simeq4.05$ Gyr). 
After that, they travelled toward us at an increasing physical speed.
Meanwhile, a possible lump in $A_0$ has evolved into structures that now reside in $A_1$ ($r=r_{PH}(t_0)=c\,\eta_0$, $\eta=\eta_0$): structures that will never be visible to us, since they fall well outside the event horizon, whose value at the present time is $r_{\scriptscriptstyle{EH}}(t_0)=c(\eta_\mathrm{max}-\eta_0)\simeq16.59\pm0.18$ Glyr.
Simultaneously with the CMBR coming from event $A_0$, we also receive the light coming from all events laying on our past light cone, for example events $C_2$ (at $r=20.0$ Glyr) and $E$ (at $r=10.0$ Glyr).
While the emission from $C_2$ was still outside the Hubble sphere, the emission from $E$ was inside and thus its photons always had negative velocities in both comoving and physical coordinates.

\noindent Studying Fig. 2 leads us to the following series of conclusions:

A. An emission of light that occurred at any space-time point inside our past light cone has reached us in every case, regardless of whether the emission point was inside or outside the Hubble sphere (see e.g. emission event $C_1$). 
In principle, we could see all emission events that lie on our past light cone. It is only a matter of telescope resolution at the various observing frequencies whether we are actually able to see them. In this regard, the JWST telescope has greatly improved in resolving power over the previous Hubble telescope, and thus it has become possible to see sources not bright enough to be detected by Hubble.

B. An emission of light that occurred at any space-time point outside our past light cone but within the event horizon, entered (or will enter) the Hubble sphere, if not already within it, and will reach us at a future time between now and infinity (see e.g. emission event $B$ at $r=30.0$ Glyr, $\eta=25.0$ Gyr that will reach us at $\eta=55.0$ Gyr or $t\simeq24.8$ Gyr). Whether this radiation can be detected will depend only on the sensitivity of the telescopes.

C. An emission of light occurring anywhere outside the event horizon will never be able to reach us. In particular, we will never see the background radiation emitted by cosmic structures having a comoving radial coordinate greater than $63.69$ Glyr. This is a relevant point that can only be seen in this diagram, since the representation in physical coordinates does not highlight this fact.

D. We can see objects that have always been receding at superluminal speeds, like the comoving structure identified with the letter $C$, of which the emission events $C_0$, $C_1$, $C_2$ are shown. In the conformal coordinate graph this is clear because the photon recession velocity is absent, whereas, as we shall see, in the physical coordinate graph this is explained because the Hubble sphere expands faster than the recession velocity of the emitted photons, catches up with them and encompasses them.

E. Once a cosmic object falls within the particle horizon, it remains within this horizon. For example, in Figure 2 the light coming from the cosmic structures placed at the radial distance of $20.0$ Glyr on the surface of last scattering ($C_0$ event of CMBR emission), entered the particle horizon at the conformal time $\eta=10.0$ Gyr and became visible to us at $\eta=20.0$ Gyr. They have been visible ever since and will still be visible in the future: e.g. the light emitted at $C_1$ ($\eta=20.0$ Gyr) reached us at $\eta=40.0$ Gyr, while the light emitted at $C_2$ on our past light cone reaches us just now.
It follows that statements referring to galaxies entering and then leaving the particle horizon are false.

F. On the contrary, there can be cosmic structures that enter and then leave the Hubble sphere. An example, shown in Fig. 2, is cosmic objects placed at a comoving radial distance of $15.0$ Glyr, which entered the Hubble sphere at $\eta\simeq31.4$ Gyr (event $D_1$) and  exited at $\eta\simeq46.0$ Gyr (event $D_2$). 
We now see these cosmic objects as they were when they crossed our past light cone, at $\eta\simeq32.1$ Gyr.

\section{Physical coordinates diagram}
\label{Sec7}
Using conformal diagrams implies a penalty: they completely hide true distances. 
The proper distance between different cosmic regions when $t\rightarrow0$ goes to zero, while in the conformal diagram it remains constant, fixed at today's value. 
The physical distances near $t=0$ are thus severely distorted, appearing much larger than they actually are, and this distortion becomes infinitely large as $t\rightarrow0$.
We therefore reproduce the key features of Figure 2 in physical coordinates (cosmic time and proper radial distance) by replacing the conformal time $\eta$ with the cosmic time $t$ and multiplying the comoving coordinates $r(t)$ at time $t$ by the scale factor $a(t)$. 
The graph takes on the appearance shown in Figure 3, where the lattice of constant comoving radial distances is translated into a series of lines starting from the origin and progressively diverging.
\begin{figure}[ht]
\includegraphics[width=.8\linewidth]{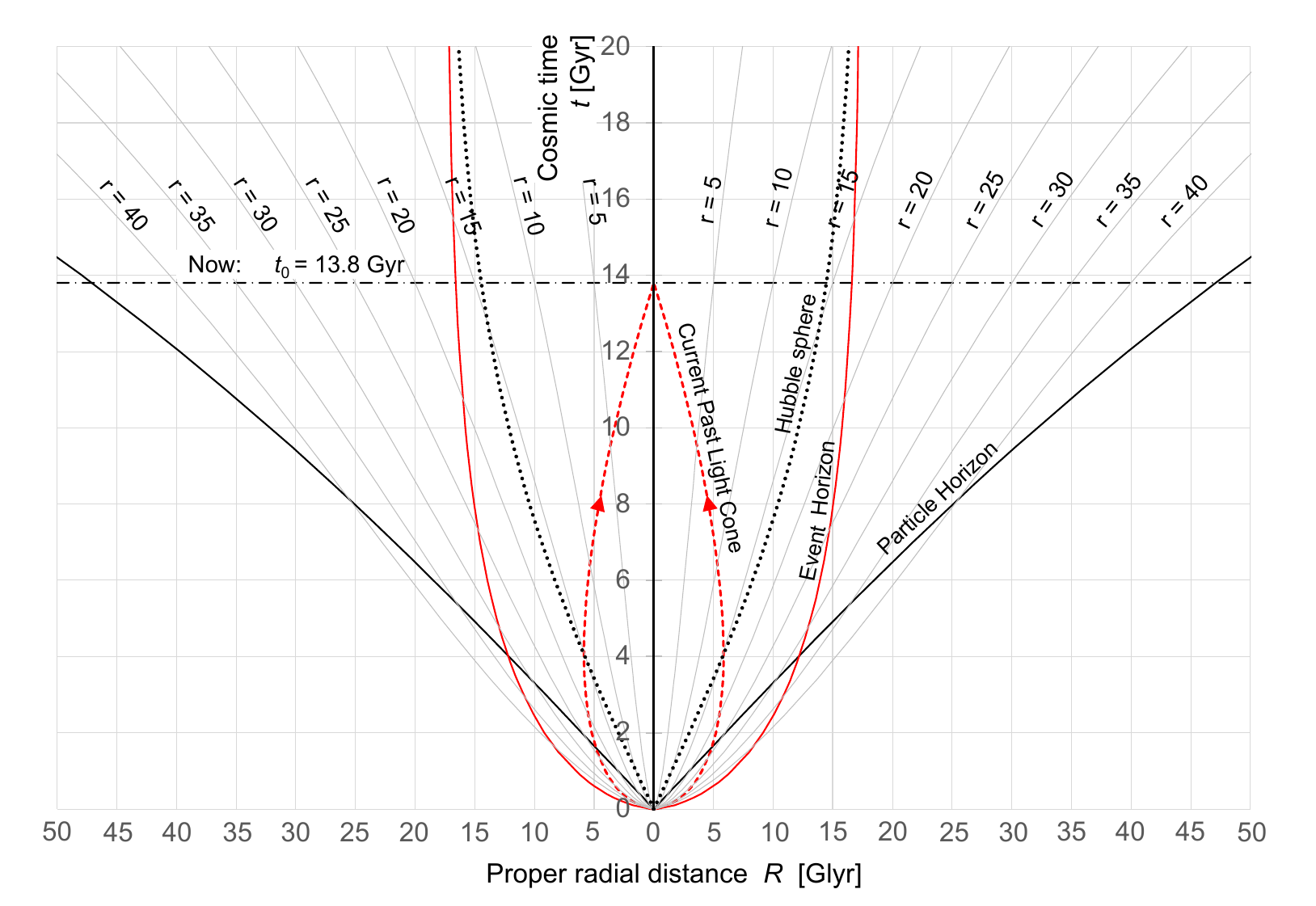}
\caption{Graph of past light cone at $t_0$, particle horizon, event horizon and Hubble sphere in physical coordinates.}
\label{fig3}
\end{figure}
The drawback of this representation, compared to the conformal one, lies in the fact that everything is ‘‘squashed’’ toward the  null proper radial distance at time zero.
For this reason Figure 4 shows a detail of this region, in which the same lengths are adopted for the units of Glyr and Gyr. 
This figure also illustrates some details of figure 2 such as light rays propagating towards the origin from the $B$ and $C_1$ emission events, and the paths of comoving particles $A$, $C$ and $D$.
\begin{figure}[ht]
\includegraphics[width=.8\linewidth]{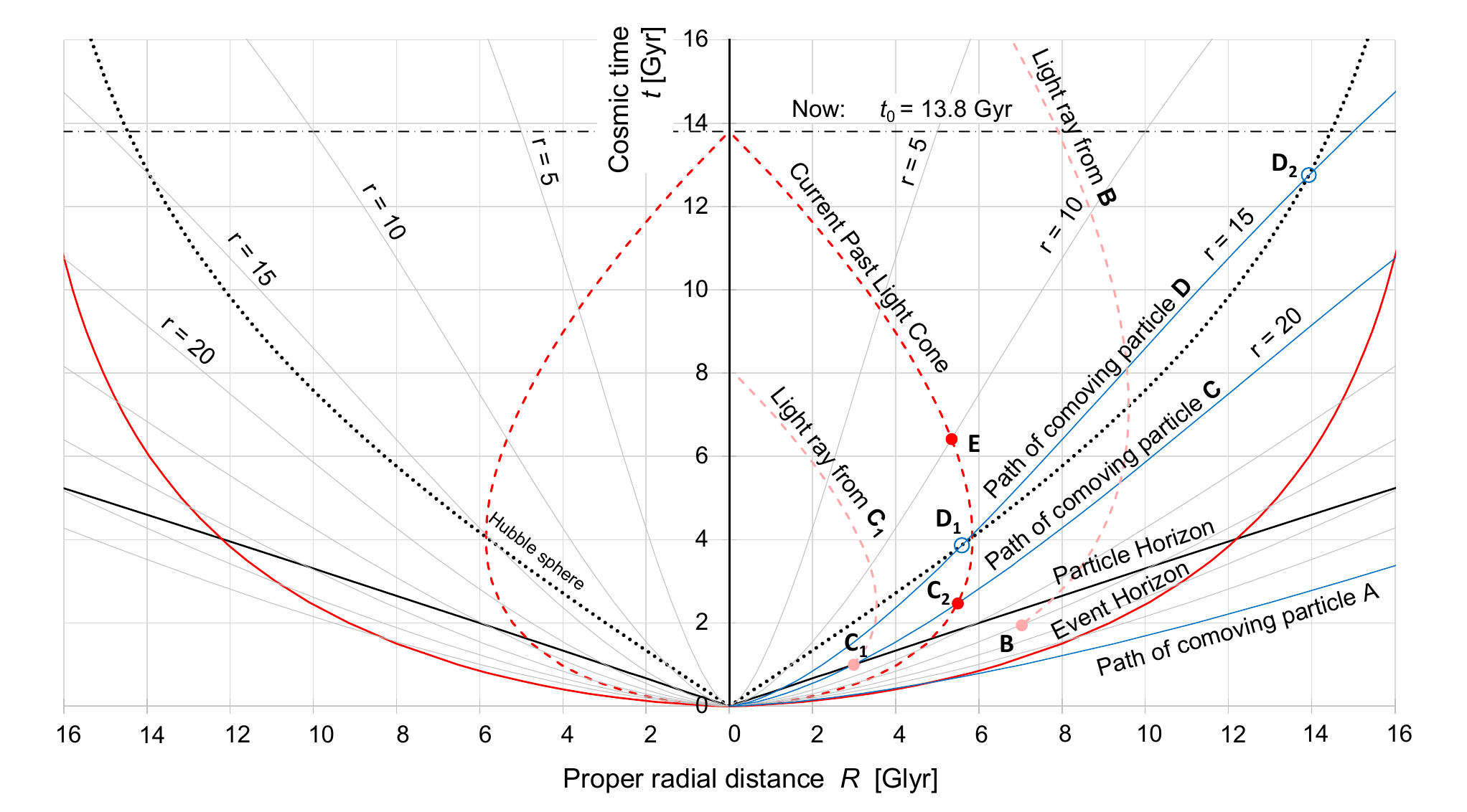}
\caption{Detail of Figure 3 with light rays emitted towards the origin at $B$ and $C_1$, and the paths of comoving particles $A$, $C$ and $D$}
\label{fig4}
\end{figure}
The functional dependencies on cosmic time of particle and event horizons expressed in terms of proper radial distance, as well as the equation of our past light cone, become more complex than those in terms of conformal coordinates (\ref{light_conf}), (\ref{particle_conf}) and (\ref{event_conf}). 
However, we can easily derive the analytical form of the different entities.

For our past light cone in terms of physical coordinates, $R_{\scriptscriptstyle{LC}}(t_0;t){\equiv}a(t)\,r_{\scriptscriptstyle{LC}}(t_0;t)$, we use the definition (\ref{light_cone}) of $r_{\scriptscriptstyle{LC}}(t_0;t)$ as well as the expression (\ref{a(t)}) for $a(t)$, obtaining:
\begin{equation}
\label{light_cone_R}
R_{\scriptscriptstyle{LC}}(t_0;t)=\frac{c}{\xi}\sinh^{2/3}\left(\xi{t}\right)\int_{\xi{t}}^{\xi{t_0}}\frac{dx}{\sinh^{2/3}(x)}{\qquad}(t\leq{t_0}).
\end{equation}
As can be seen from Figures 3 and 4, the past light cone takes on the appearance of a ‘‘drop’’ because toward time zero the scale factor tends to zero. The photons emitted at time zero were originally very close to the origin, from which they moved away until they crossed the Hubble sphere at $t\simeq4.06$ Gyr reaching the maximum proper radial distance, $R_{\scriptscriptstyle{LC}}(t_0;4.06)\simeq5.85$ Glyr. They then began to approach us. 
Like Figure 2, Figure 4 also shows the light tracks toward us for the emissions that occurred at the $C_1$ and $B$ events.
These paths belong to past light cones pertaining to cosmic times $t^*\simeq8.12$ Gyr and $t^*\simeq24.85$ Gyr, respectively. Like our past light cone, all these light cones reach their maximum radial proper distance when they cross the Hubble sphere.

For the particle horizon in physical coordinates, $R_{\scriptscriptstyle{PH}}(t){\equiv}a(t)\,r_{\scriptscriptstyle{PH}}(t)$, we use the definition (\ref{particle_horizon}) of $r_{\scriptscriptstyle{PH}}(t)$ obtaining:
\begin{equation}
\label{Particle_horizon_R}
R_{\scriptscriptstyle{PH}}(t)=\frac{c}{\xi}\sinh^{2/3}\left(\xi{t}\right)\int_{0}^{\xi{t}}\frac{dx}{\sinh^{2/3}(x)}.
\end{equation}
Presently $R_{\scriptscriptstyle{PH}}(t_0)=r_{\scriptscriptstyle{PH}}(\eta_0)=c\,\eta_0=47.10\pm0.39$ Glyr, a distance corresponding to $3.41$ times the age of the universe, which means that comoving cosmic structures belonging to that region (region $A$) travelled away from us at an effective speed larger than three times the speed of light. Currently the recession speed of these cosmic structures, on the basis of Eq. (\ref{v_rec}), is $v_{\scriptscriptstyle\mathrm{rec}}=R_{\scriptscriptstyle{PH}}(t_0)H(t_0)=(3.257\pm0.033)\,c$.
Although $\lim_{t\to\infty} r_{\scriptscriptstyle{PH}}(t)=63.69\pm0.49$ Glyr, in physical coordinates $\lim_{t\to\infty}R_{\scriptscriptstyle{PH}}(t)=\infty$. This implies that the point in space that emitted the background radiation that will reach us in an infinitely large time will be at an infinitely large distance.
However, as we already pointed out, this does not mean that we will be able to see the entire universe: regions that are at comovong coordinates $r>63.69$ Glyr will never be visible to us.

For the Hubble radius in physical coordinates, $R_{\scriptscriptstyle{H}}(t){\equiv}a(t)\,r_{\scriptscriptstyle{H}}(t)=c/H(t)$, we use the definition (\ref{Hubble}) for $H(t)$ and obtain:
\begin{equation}
\label{Hubble_sphere_R}
R_{\scriptscriptstyle{H}}(t)=\frac{3c}{2\xi}\tanh{\left(\xi{t}\right)}.
\end{equation}
Its value today is $R_{\scriptscriptstyle{H}}(t_0)=c/H_0=14.46\pm0.09$ Glyr. Furthermore,
\begin{equation}
\label{RH_Limit}
R_{\scriptscriptstyle{H\infty}}\equiv\lim_{t\to\infty} R_{\scriptscriptstyle{H}}(t)=\frac{3c}{2\xi}=17.42\pm0.13\;\mathrm{Glyr}.
\end{equation}
This means that:
\begin{equation}
H_\infty\equiv\lim_{t\to\infty}H(t)=\frac{2}{3}\xi=H_0\sqrt{\Omega_{\Lambda}}=56.16\pm0.42\;\mathrm{(km/s)/Mpc}
\end{equation}
while
\begin{equation}
H(t)=H_0\sqrt{\Omega_{\Lambda}}\,\coth(\xi{t}).
\end{equation}

Finally, for the event horizon in physical coordinates, $R_{\scriptscriptstyle{EH}}(t){\equiv}a(t)\,r_{\scriptscriptstyle{EH}}(t)$, we use the definition (\ref{event_horizon}) of $r_{\scriptscriptstyle{EH}}(t)$ obtaining
\begin{equation}
\label{Event_horizon_R}
R_{\scriptscriptstyle{EH}}(t)=\frac{c}{\xi}\sinh^{2/3}\left(\xi{t}\right)\int_{\xi{t}}^{\infty}\frac{dx}{\sinh^{2/3}(x)}.
\end{equation}
We have seen that, at all times, the event horizon is always larger than the Hubble radius and that when cosmic time tends to infinity, it tends asymptotically to the Hubble radius. We have anticipated that this feature is due to the fact that the physical velocity of photons proceeding along the event horizon tends progressively to zero.
We can now prove this assertion by taking the derivative with respect to cosmic time of the expression (\ref{Event_horizon_R}):
\begin{equation}
\label{v_rec_EH_1}
\frac{dR_{\scriptscriptstyle{EH}}}{dt}=c\,\frac{2}{3}\sinh^{-1/3}\left(\xi{t}\right)\cosh\left(\xi{t}\right)\int_{\xi{t}}^{\infty}\frac{dx}{\sinh^{2/3}(x)}-c.
\end{equation}
For large values of $\xi{t}$ we can substitute $\exp(\xi{t})/2$ for $\sinh(\xi{t})$ and $\cosh(\xi{t})$, thus obtaining
\begin{equation}
\label{v_rec_EH_2}
\frac{dR_{\scriptscriptstyle{EH}}}{dt}=c\frac{e^{\frac{2}{3}\xi{t}}}{3\cdot2^{-\frac{1}{3}}}\int_{\xi{t}}^{\infty}2^\frac{2}{3}e^{-\frac{2}{3}x}dx-c= c-c=0.
\end{equation}
We can also verify that the asymptotic value reached by $R_{\scriptscriptstyle{EH}}$ coincides with the value found for $R_{\scriptscriptstyle{H\infty}}$:
\begin{equation}
\label{Event_horizon_Limit}
R_{\scriptscriptstyle{EH\infty}}\equiv\lim_{t\to\infty} R_{\scriptscriptstyle{EH}}(t)=\frac{c}{\xi}\lim_{t\to\infty}\frac{\frac{d}{dt}\left[\int_{\scriptscriptstyle\xi{t}}^{\scriptscriptstyle\infty}\frac{dx}{\sinh^{2/3}(x)}\right]}{\frac{d}{dt}\left[\frac{1}{\sinh^{2/3}\left(\xi{t}\right)}\right]}=\frac{3c}{2\xi}.
\end{equation}
While for $t\rightarrow\infty$ the comoving Hubble radius $r_{\scriptscriptstyle{H}}$ and the comoving event horizon $r_{\scriptscriptstyle{EH}}$ tend to $r=0$, in terms of physical radial distance $R_{\scriptscriptstyle{H}}$ and $R_{\scriptscriptstyle{EH}}$ are always increasing in time and tend to the common constant limit (\ref{RH_Limit}).
The event horizon today is $R_{\scriptscriptstyle{EH}}(\eta_0)=r_{\scriptscriptstyle{EH}}(t_0)=16.59\pm0.18$ Glyr: radiation emitted today by sources that are within this maximum radial distance will always be able to reach us in the future.

\begin{figure}[ht]
\includegraphics[width=.8\linewidth]{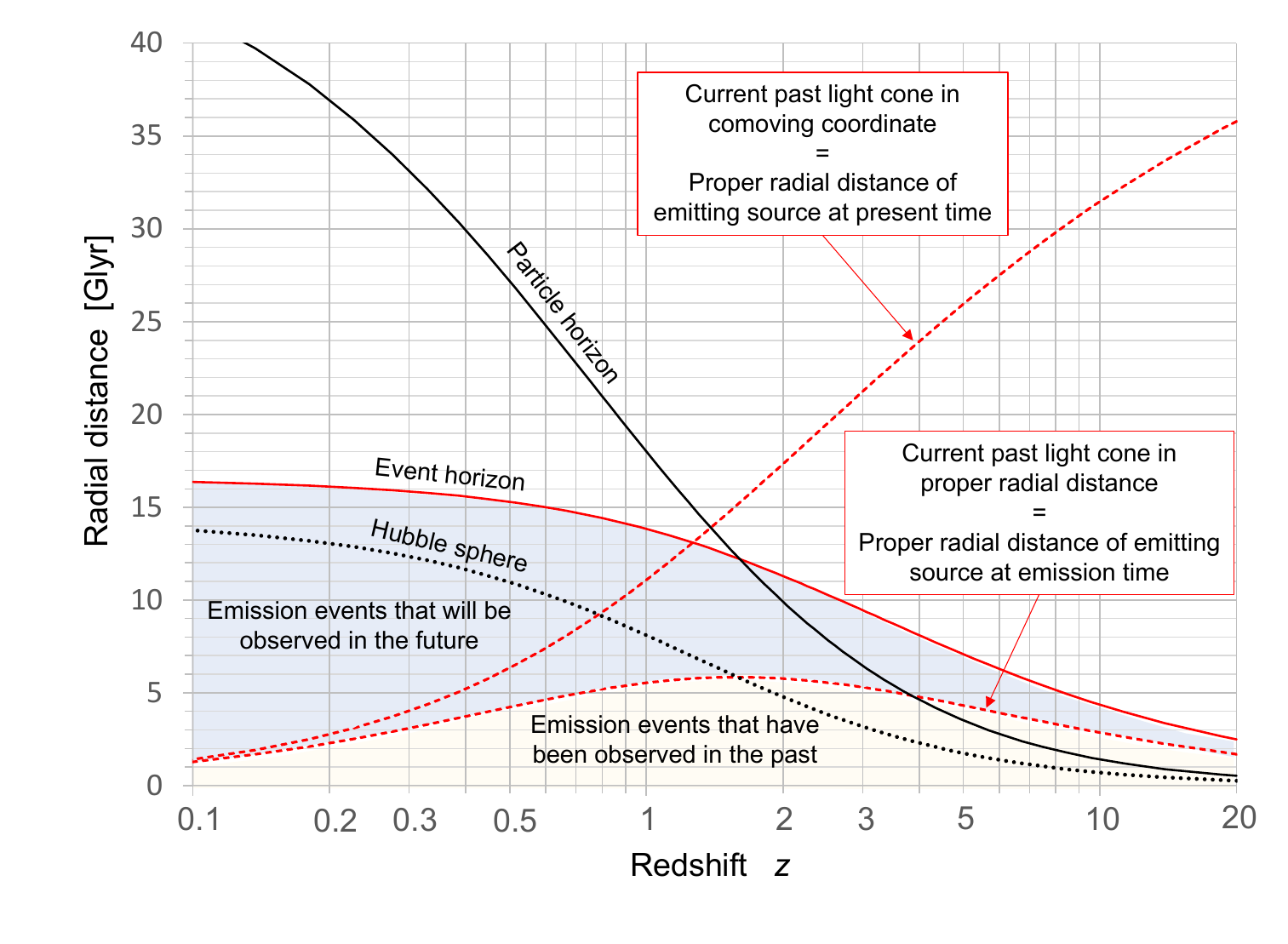}
\caption{Light cone, particle horizon, event horizon and Hubble sphere as a function of the redshift of incoming radiation. The regions where emission events that could be observed in the past and can be observed in the future are indicated. The boundary between the two regions is given by our past light cone.}
\label{fig5}
\end{figure}
\section{Horizons and Redshift}
\label{Sec8}

The connection of the previously analysed quantities with experimental observables is made through the cosmological redshift $z$. 
The redshift allows the direct determination of the scale factor by means of Eq. (\ref{eq:redshift}) and, through equation (\ref{a(t)}), the $\Lambda$CDM model provides an evaluation of the cosmic time at which the emission event occurred.
By virtue of the relationship between redshift and emission time, all quantities can be described and represented as a function of redshift.
Figure 5 gives such a representation.
In particular, the equation for the past light cone in terms of physical coordinates, Eq. (\ref{light_cone_R}), gives the proper radial distance of the emitting body at the time of emission.
On this curve lie all the emission events that we can detect today. 
Below it lie the emission events whose radiation has come to us in the past (pale yellow area in Figure 5), while above, but below the event horizon curve, are the emission events whose radiation will reach us in the future (pale blue area in Figure 5).
The equation for the past light cone (\ref{light_cone}) gives the comoving radial distance of the emitting body, which coincides with its current proper radial distance.

\section{Conclusions}
Referring to the so-called standard cosmological $\Lambda$CDM model, the properties of cosmological horizons (particle horizon, optical horizon, event horizon), the past light cone at various epochs (especially our past light cone) and the Hubble sphere have been illustrated. 
With the help of graphical representations, both in conformal and physical coordinates, the radial motion of photons was examined and their physical velocities with respect to us were derived.
It has thus been clarified what portions of the universe we can and will be able to see in the future, and what we can never see.
The treatment is aimed at teachers who wish to comprehensively explain cosmological horizons in their Relativity or Cosmology courses.
%

%


\begin{thebibliography}{99}
%
\bibitem{2022 Crane}
Crane L., ‘‘James Webb Space Telescope releases dazzling first science images’’, NewScientist, Space, 12 July 2022
%
\bibitem{1956 Rindler}
Rindler, W. ‘‘Visual Horizons in World-Models’’, Mon. Not. Roy. Astr. Soc., \textbf{116}, 662 (1956)
%
\bibitem{1991 Harrison}
Harrison E., ‘‘Hubble Sphere and Particle Horizons’’, Astrophys. J., \textbf{383}, 60-65 (1991)
%
\bibitem{1993 Ellis}
Ellis G. F. R. and Rothman T., ‘‘Lost horizons‘‘, Am. J. Phys. \textbf{61} (10), 883-893 (1993)
%
\bibitem{2012 Margalef}
Margalef-Bentabol B., Margalef-Bentabolb J. and Cepa J., ‘‘Evolution of the cosmological horizons in a concordance universe’’, J. Cosmol. Astropart. Phys. \textbf{12}, 035 (2012) 
%
\bibitem{2013 Margalef}
Margalef-Bentabol B., Margalef-Bentabolb J. and Cepa J., ‘‘Evolution of the cosmological horizons in a concordance universe’’, J. Cosmol. Astropart. Phys. \textbf{02}, 015 (2013) 
%
\bibitem{2015 Faraoni}
Faraoni V. ‘‘Cosmological and Black Hole Apparent Horizons’’, Lecture Notes in Physics \textbf{907}, Springer (2015)
%
\bibitem{2003 Davis}
Davis T. M., ‘‘Fundamental Aspects of the Expansion of the Universe and Cosmic Horizons’’, PhD Thesis, University of New South Wales, Sydney (2003)
%
\bibitem{2004 Davis}
Davis T. M. and Lineweaver C. H., ‘‘Expanding Confusion: Common Misconceptions of Cosmological Horizons and the Superluminal Expansion of the Universe’’, Publ. Astron. Soc. Aust., \textbf{21} 97-109 (2004)
%
\bibitem{2005 Lineweaver}
Lineweaver C. H. and Davis T. M., ‘‘Misconceptions about the Big Bang’’, Sci. Am., March 2005, 37-45
%
\bibitem{2020 Planck Collaboration}
Planck Collaboration, ‘‘Planck 2018 results. VI. Cosmological parameters’’, Astron. Astrophys. \textbf{641} A6 (2020) - p.40 and rightmost column of Table 2 on p. 15
%
\bibitem{2019 Carroll}
Carroll S. M., \textit{Spacetime and Geometry - An Introduction to General Relativity}, Cambridge University Press, (2019) - pp. 332-334
%
\bibitem{2002 Coles}
Coles P. and Lucchin F., \textit{Cosmology - The Origin and Evolution of Cosmic Structure}, John Wiley \& Sons, Ltd (2002) - p. 16 and p. 30
%
\bibitem{1929 Hubble}
Hubble, E., "A relation between distance and radial velocity among extra-galactic nebulae". Proc. Natl. Acad. Sci. \textbf{15} (3) 168–73 (1929)
%
\bibitem{uncertainties}
Observe that by virtue of Eq. (\ref{Closure_2}) the uncertainties of $\Omega_m$ and $\Omega_{\Lambda}$ are equal.
%
\bibitem{hypergeometric}
For a review of Gaussian hypergeometric functions, see e.g. \\ https://mathworld.wolfram.com/HypergeometricFunction.html
%
\bibitem{nintegration}
An educational presentation on numerical integration can be found at \\ https://mathworld.wolfram.com/NumericalIntegration.html.

\end{thebibliography}
\end{document}